\documentclass[11pt,a4paper,psamsfonts]{amsart}
\usepackage{amsfonts,amscd,amsmath,amssymb,amsthm}
\usepackage{times}
\newtheorem{lem}{Lemma}[section]
\newtheorem{theo}{Theorem}[section]
\newtheorem{defi}{Definition}[section]
\newtheorem{cor}{Corollary}[section]

\newtheorem{re}{Remark}[section]

\DeclareMathOperator{\tr}{Tr}
\DeclareMathOperator{\Ran}{Ran}

\title{Perron-Frobenius theory for positive maps on trace
  ideals}
\author[R. Schrader]{R. Schrader$^\ast$}

\thanks{$^\ast$ e-mail: schrader@physik.fu-berlin.de,\\
Supported in part by DFG SFB 288 ``Differentialgeometrie und Quantenphysik''}

\address{Institut f\"{u}r
Theoretische Physik\\ Freie Universit\"{a}t Berlin, Arnimallee 14\\ D-14195 Berlin, Germany}

\email{schrader@physik.fu-berlin.de}
\begin{document}

\begin{abstract}
This article provides sufficient conditions for positive maps on
the Schatten classes $\mathcal J_{p},\;1\le p<\infty$ of bounded operators on a
separable Hilbert space such that a corresponding Perron-Frobenius theorem
holds. With applications in quantum information theory in mind sufficient
conditions are given for a trace
preserving, positive map on $\mathcal J_{1}$, the space of trace
class operators, to have a unique, strictly positive density matrix
which is left invariant under the map. Conversely to any given strictly
positive density matrix there are trace preserving, positive maps
for which the density matrix is the unique Perron-Frobenius vector.
\end{abstract}

\maketitle 
\centerline{\em Dedicated to S. Doplicher and J.E. Roberts on the
  occasion of their 60th birthday}


\section{Introduction}
In the theory of quantum information the transmission through noisy
channels plays an important role. Usually it is described by what
physicists either call a quantum operation (see e.g. \cite{Nielsen}) or
a stochastic
map (\cite{AU}, see also \cite{KiRu}) or a super-operator (see
e.g. \cite{Preskill}) and what mathematicians call a completely
positive (trace preserving) map. In mathematics positive maps were first
studied by Kadison in the context of $C^{\star}$-algebras
\cite{Kadison} and completely positive maps by Stinespring \cite{St}. 
In quantum physics it was first
introduced by Haag and Kastler, who called it an operation \cite{HaK}. They
were then studied in more detail in \cite{HeK1}. \cite{KK1,KK}
contains an extensive discussion, how this concept naturally arises in
quantum physics.

Because they preserve positivity such maps naturally fit into a 
context where it makes sense to ask about a formulation of a corresponding
Perron-Frobenius theory.

As the name indicates, the earliest results relating positivity of a finite
dimensional linear
map to the non-degeneracy of an eigenvalue and the positivity of the
resulting (Perron-Frobenius) eigenvector is due to Perron and Frobenius
\cite{Perron,Frobenius1,Frobenius2}. The first extension to the
infinite dimensional was given by Jentzsch \cite{Jentzsch}. Since then
there has been an extensive development (see
e.g. \cite{Ando,KreinRutman,Sch}).

The interest in quantum physics and quantum field theory originates
from the observation, that
very often the ground state of a quantum system with Hamilton operator
$H\ge 0$ is non-degenerate and nowhere vanishing. Indeed, under
suitable circumstances such a ground
state can be viewed as the Perron-Frobenius vector for $\exp -tH,
t>0$, an observation first made by Glimm and Jaffe \cite{GJ} (see
also \cite{RS,GlimmJaffe} for an overview and with references
to other articles).

The Perron-Frobenius theory and related topics for positive linear
maps on linear spaces of operators  was first studied in
\cite{EH,WaEn,Groh,AH}, where
in the first article the space of operators was the $C^{\star}$-algebra
of all linear operators on a finite dimensional Hilbert space. In the
second and third articles the discussion was extended to the infinite
dimensional case  and in the fourth article the analysis was carried
out in the context of a von Neumann algebra.

The present article can be viewed as an extension of the discussion in
\cite{EH} to the infinite dimensional case. Indeed, we will consider
positive maps on the Schatten classes $\mathcal J_{p},\;1\le p\le\infty$ and
provide a Perron-Frobenius theory on such spaces. As in the usual
context the notions of
ergodicity and positivity improving will play an essential role.
Since density matrices are elements of $\mathcal J_{1}$ our main focus will
be on the case $p=1$. In particular we will
show that any compact, ergodic, completely positive and trace preserving map in
$\mathcal J_{1}$ has a unique density matrix invariant under this map
(i.e. the eigenvalue equals 1) and which therefore can be viewed as
the Perron-Frobenius vector for this map.

The article is organized as follows. In Section 2 we will briefly
review the concepts needed for our discussion. In Section 3 we will
prove a Perron-Frobenius theorem (non-degeneracy of a certain
eigenvalue), provided such an eigenvalue exists, a typical condition
needed in infinite dimensional contexts. Section 4 will provide sufficient
conditions for such an eigenvalue to exist. In Section 5 we will give
examples. Also we will prove a converse in the sense that to any
density matrix $>0$ there are completely positive maps for which this
density matrix is the Perron-Frobenius eigenvector.

\section{Positive and Completely Positive Maps}
We start by recalling and establishing some definitions and facts related to
trace ideals and to the concept of positive and completely positive maps
( for more details on trace ideals see e.g.
\cite{Schatten,Simon1} and on positive and completely positive maps e.g.
\cite{St,Stoermer,Choi,Davies,Ta,Paulsen}). Let $\mathcal H$ be any
complex, separable
Hilbert space with scalar product denoted by $\langle\;,\rangle$. 
By $\mathcal B(\mathcal H)$ we denote
the $C^{\star}$-algebra of all bounded operators on $\mathcal H$ equipped
with the norm $||\;||$. By $\mathbb{I}\in\mathcal B(\mathcal H)$ we denote the identity map
on $\mathcal H$.

For any
$A\in \mathcal B(\mathcal H)$ we set $|A|=(A^{\star}A)^{1/2}\in\mathcal B(\mathcal H)$ where
$^{\star}$ denotes the adjoint. Also we let $\tr$ be the trace
operation on $\mathcal H$. For $1\le p<\infty$ let
$\mathcal J_{p}=\mathcal J_{p}(\mathcal H)$ be
the Schatten class consisting of all elements $A\in \mathcal B(\mathcal H)$
such that $|A|^{p}$ is trace class. $\mathcal J_{p}$ is equipped with the norm
$||A||_{p}=(\tr(|A|^{p}))^{1/p}$ making each $\mathcal J_{p}$ a complete
Banach space. One has $A^{\star}\in\mathcal J_{p}$ if $A\in\mathcal J_{p}$, i.e. the
space $\mathcal J_{p}$ is self-adjoint in $\mathcal B(\mathcal H)$, such that
$||A^{\star}||_{p}=||A||_{p}$. Also we
set $\mathcal J_{\infty}=\mathcal B(\mathcal H)$ again
with the same norm, i.e. $||\;||_{\infty}=||\;||$. We will frequently
make use of the trivial identity $||A||_{p}=||\,|A|\,||_{p}$ valid for
all $A\in\mathcal J_{p}$ and all $1\le p\le\infty$.
Via the trace $\mathcal J_{q}$ is the dual of
$\mathcal J_{p}$ with $1/p+1/q=1$ for all $1\le p <\infty$, i. e. each
continuous linear functional on $\mathcal J_{p}$ is given by an $A\in\mathcal J_{q}$
in the the form $\tr(A^{\star}B),\,B\in\mathcal J_{p}$. In particular the
spaces $\mathcal J_{p}$ are reflexive for all $1<p<\infty$.
We prefer to use the adjoint in describing linear functionals
since for $p=2$,
in which case $\mathcal J_{2}$ is the space of Hilbert-Schmidt operators on $\mathcal H$,
the scalar product making it a Hilbert space is just given by
$\langle A,B\rangle_{2}=\tr(A^{\star}B)$. Also one has the
H\"{o}lder inequality $|\tr(A^{\star}B)|\le ||A||_{q}||B||_{p}$.
Let $\mathcal C=\mathcal C(\mathcal H)=\mathcal C_\infty$ be the closed positive cone of all
elements $A\ge 0$ in $\mathcal B(\mathcal H)$. As usual we write $A\ge 0$
and $A>0$ if the relations $\langle\varphi,A\varphi\rangle\,\ge\, 0$ 
and $\langle\varphi,A\varphi\rangle\,>\,0$
respectively hold for all $0\neq \varphi\in \mathcal H$. Correspondingly we
write $A>B$ (or $B<A$) and $A\ge B$ (or $B\le A$) if $A-B>0$ and
$A-B\ge 0$ respectively. Obviously $A\ge B>0$ or $A>B\ge 0$ implies
$A>0$. As is common, $A$ is said to be positive definite if $A>0$.
Set $\mathcal C_{p}=\mathcal C_{p}(\mathcal H)=\mathcal C\cap\mathcal J_{p}$, a closed set in $\mathcal J_{p}$.
Also $A\ge 0$ in $\mathcal J_{p}$ if and only if $\tr(BA)\ge 0$ for all
$0\le B\in\mathcal J_{q}$. Correspondingly $A>0$ in $\mathcal J_{p}$ if and only if
$\tr(BA)>0$ for all $0\le B\neq 0$ in
$\mathcal J_{q}\;(1/p+1/q=1,\,1\le p\le\infty)$. In this sense the cones
$\mathcal C_{p}$ and $\mathcal C_{q}$ are dual to each other.
The closed set $\mathcal C_{1,1}=\{A\in \mathcal C_{1}|\tr A=||A||_{1}=1\}$ in $\mathcal J_{1}$ is
the set of all density matrices in $\mathcal H$.

By definition a positive map $\phi$ in $\mathcal J_{p}$ is a linear map from $\mathcal J_{1}$
into itself, which leaves $\mathcal C_{p}$ invariant. $\phi$ is called
$n$-positive if the induced map $\phi_{n}=\phi\otimes\mathbb{I}_{n}$ in
$\mathcal J_{p}\otimes \mathcal B(\mathcal H_{n})$ also leaves the corresponding cone
$\mathcal C_{p}(\mathcal H\otimes\mathcal H_{n})$ of non-negative elements invariant.
Here $\mathcal H_{n}$ is any Hilbert space of dimension $1\le n<\infty$.
Obviously if $\phi$ is $n$-positive, then it is also
$n^{\prime}$-positive for any $1\le n^{\prime}\le n$ as is
$\lambda\phi,\, \lambda>0$. If $\phi$ and
$\phi^{\prime}$ are both $n$-positive on $\mathcal J_{p}$ then so is their
composite $\phi\circ\phi^{\prime}$ and their sum
$\phi+\phi^{\prime}$. So the $n$-positive maps in $\mathcal J_{p}$
form a cone in the linear space of all linear maps in $\mathcal J_{p}$.

If $\phi$ is $n$-positive for all $n$, then $\phi$ is called
completely positive. Thus the map $\phi_{\alpha}:
A\rightarrow\alpha A\alpha^{\star}$ for any $\alpha\in\mathcal B(\mathcal H)$ is completely
positive on $\mathcal J_{p}$ for all $1\le p\le\infty$ as is any finite
linear combination
$\phi_{\underline{\alpha}}=\sum_{i}\phi_{\alpha_{i}}$ with
$\alpha_{i}\in\mathcal B(\mathcal H)$. If $\alpha$ has an inverse $\alpha^{-1}$ in $\mathcal B(\mathcal H)$
then $\phi_{\alpha}^{-1}=\phi_{\alpha^{-1}}$. Also
$\phi_{\alpha}\circ\phi_{\alpha^{\prime}}=\phi_{\alpha\,\alpha^{\prime}}$
and $\phi_{\lambda\alpha}=|\lambda|^{2}\phi_{\alpha}$ for
$\alpha,\,\alpha^{\prime}\in\mathcal B(\mathcal H)$ and $\lambda\in\mathbb{C}$.
As it turns
out at least for $p=1$ the induced
linear map $\phi\otimes\mathbb{I}_{\mathcal H^{\prime}}$ from
$\mathcal J_{p=1}\otimes\mathcal B(\mathcal H^{\prime})$ then also leaves the corresponding
cone invariant for any separable Hilbert space $\mathcal H^{\prime}$ (see
e.g. \cite{KK}). Since
density matrices are in $\mathcal J_{1}$, the case $p=1$ is of most interest
in quantum physics. Then there are good physical reasons to consider
completely positive
maps rather than only positive maps (see e.g, \cite{KK}).

Although the following lemma is well known in similar contexts, we
still will provide the short proof.
\begin{lem}
The following relation holds if the map $\phi$ in
$\mathcal J_{p}\;(1\le p\le\infty)$ is positive.
\begin{equation}
\label{star3}
\phi(A)^{\star}=\phi(A^{\star}),\, A\in \mathcal J_{p}.
\end{equation}
\end{lem}
\begin{proof} We first consider the case when $A$ is self-adjoint,
i.e. $A^{\star}=A$.
Write $A=A_{+}-A_{-}$ with $A_{\pm}\in\mathcal C_{p},\,A_{+}A_{-}=A_{-}A_{+}=0$
and $|A|=A_{+}+A_{-}$. Here $\pm A_{\pm}$ are the positive and negative
parts of $A$, obtainable from the spectral representation of $A$ or
more explicitly as $A_{\pm}=1/2(|A|\pm A)$. Obviously
$||A_{\pm}||_{p}\le ||A||_{p}$.
Since $\phi(A_{\pm})\ge 0$,
$\phi(A)=\phi(A_{+})-\phi(A_{-})$ is self-adjoint.
For arbitrary $A\in\mathcal B(\mathcal H)$ write
$A=\Re A+i\Im A\in \mathcal B(\mathcal H)$ with
$\Re A=1/2(A+A^{\star}),\,\Im A=1/2i(A-A^{\star})$, such that
$\Re A,\,\Im A\in\mathcal B(\mathcal H)$ are self-adjoint with
$\Re A,\,\Im
A\in\mathcal J_{p}$ whenever $A\in\mathcal J_{p}$. More precisely we have the a priori
bound $||\Re A||_{p}\le ||A||_{p},\,
||\Im A||_{p}\le ||A||_{p}$. This gives the decomposition
\begin{equation*}
A=(\Re A)_{+}-(\Re A)_{-}+i((\Im A)_{+}-(\Im A)_{-})
\end{equation*}
and hence by the linearity of $\phi$
\begin{equation}
\label{decomp}
\phi(A)=\phi((\Re A)_{+})-\phi((\Re A)_{-})+i\phi((\Im A)_{+})
-i\phi((\Im A)_{-})
\end{equation}
with
$||\phi(A)||_{p}\le ||\phi((\Re A)_{+})||_{p}+||\phi((\Re A)_{-})||_{p}
+||\phi((\Im A)_{+})||_{p}+||\phi((\Im A)_{-})||_{p}$. In particular
(\ref{decomp}) shows that indeed (\ref{star3}) holds.
\end{proof}
We also note the following. The relation
$-|A|\le A\le|A|$ for $A=A^{\star}\in\mathcal J_{p}$ implies
$-\phi(|A|)\le\phi(A)\le\phi(|A|)$, whenever the map $\phi$ in
$\mathcal J_{p}$ is positive. This in turn gives
\begin{equation}
\label{mon1}
|\phi(A)|\le\phi(|A|)
\end{equation}
valid for any $A=A^{\star}\in\mathcal J_{p}$ and any positive map $\phi$ in
$\mathcal J_{p}$. We do not know whether (or when) this relation continues to
hold when the condition $A=A^{\star}$ is dropped (see, however,
(\ref{bound5a}) below for a weaker result, which will suffice for our
purposes).
Observe by comparison that for any $n\times n$ matrix $S$ with
non-negative entries, the classical context for the Perron-Frobenius
theorem, one has
$|S\underline{z}|\le S|\underline{z}|$ for $\underline{z}\in
\mathbb{C}^{n}$. Here $|\underline{z}|$ is the vector whose components are the
absolute values of the corresponding components of $\underline{z}$.
Also for real vectors $\underline{x}=\{x_{i}\}$ and
$\underline{y}=\{y_{i}\}$ by definition $\underline{x}\le
\underline{y}$ if and only if $x_{i}\le y_{i}$ holds for all $i$.

Furthermore one does not necessarily have
$\phi(A)_{\pm}=\phi(A_{\pm})$ for any
$A=A^{\star}\in\mathcal J_{p}$. However, since $-A_{-}\le A\le A_{+}$ implies
$-\phi(A_{-})\le \phi(A)\le\phi(A_{+})$, the
inequalities $\phi(A)_{\pm}\le\phi(A_{\pm})$ hold.

\begin{lem} Any positive map $\phi$ in $\mathcal J_{p}\;(1\le p\le\infty)$ is
continuous, i.e. it satisfies $||\phi||_{p}<\infty$.
\end{lem}
Here we denote by $||\phi||_{p}$ the norm of any continuous linear map
$\phi$ in $\mathcal J_{p}$, such that
\begin{equation}
||\phi||_{p}=\sup_{||A||_{p}\le 1}||\phi(A)||_{p}.
\end{equation}
In particular
\begin{equation}
\label{bound02}
||\phi||_{1}=\sup_{A\in \mathcal C_{1,1}} \tr(\phi(A)),
\end{equation}
if $\phi$ is completely positive on $\mathcal J_{1}$ (see e.g. \cite{Davies,KK}).

\begin{proof} We adapt a
standard proof (see e.g. \cite{Paulsen}, p. 19) used in the context of
positive maps on non-unital $C^{\star}$-algebras. We first claim that it
suffices to prove boundedness on $\mathcal C_{p}$. Indeed, this follows from the
decomposition (\ref{decomp}) and the related bounds.
Assume that $\phi$ is not bounded on $\mathcal C_{p}$. Then there are
$A_{n}\in\mathcal C_{p}$ with $||A_{n}||_{p}\le 1$ and
$||\phi(A_{n})||_{p}\ge n^{3}$. Let
$A=\sum_{n}1/n^{2}A_{n}\in\mathcal C_{p}$, such that $0\le 1/n^{2}A_{n}\le A$
holds for all $n$.

We need the fact that
\begin{equation}
\label{mon}
||B||_{p}\le ||B^{\prime}||_{p}
\end{equation}
holds for any $0\le B\le B^{\prime}\in\mathcal C_{p}$. The case $p=\infty$ is
trivial. When $1\le p \le\infty$ take
$\varphi_{n}$ to be a complete orthonormal basis in $\mathcal H$ diagonalizing
$B$. Then we have
\begin{eqnarray*}
||B||_{p}^{p}&=&\sum_{n}\langle\varphi_{n},B\varphi_{n}\rangle^{p}
    \;\le\;\sum_{n}\langle\varphi_{n},B^{\prime}\varphi_{n}\rangle^{p}\\
      &\le&\sum_{n}\langle\varphi_{n},B^{\prime^{p}}\varphi_{n}\rangle
             \;=\; ||B^{\prime}||_{p}^{p},
\end{eqnarray*}
proving (\ref{mon}).
Here we have used the estimate
\begin{equation}
\label{pos40}
|\langle\varphi,A\varphi\rangle|^{p}\le \langle\varphi,|A|^{p}\varphi\rangle,
\end{equation}
valid for any $A=A^{\star}\in \mathcal B(\mathcal H)$, any normalized $\varphi\in\mathcal H$ and
any $1\le p$ (see e.g. \cite{Simon1}, p.21).
Since $\phi$ is positive (\ref{mon}) gives
$n\le 1/n^{2}||\phi(A_{n})||_{p}\le ||\phi(A)||_{p}$, which is a
contradiction.
\end{proof}
For any continuous linear map $\phi$ in
$\mathcal J_{p},\,1\le p <\infty$ let $\phi^{\star}$ be the ``adjoint'' continuous
linear map from $\mathcal J_{q},\,1/q+1/p=1$ into itself given by the
relation $\tr((\phi^{\star}(A))^{\star}B)=\tr(A^{\star}\phi(B))$ for all
$A\in \mathcal J_{q}$ and $B\in\mathcal J_{p}$. In particular
$\phi^{\star\star}=\phi$ holds whenever $1<p<\infty$. Also
$\phi_{\underline{\alpha}}^{\star}=\phi_{\underline{\alpha}^{\star}}$
with
$\underline{\alpha}^{\star}=(\alpha_{1}^{\star},...\alpha_{i}^{\star},...)$
by the cyclicity of the trace.

By definition a continuous linear map $\phi$ in $\mathcal J_{1}$ is called trace
preserving if
$\tr(\phi(A))=\tr(A)$ holds for all $A\in\mathcal J_{1}$ and this in turn is
equivalent to the relation $\phi^{\star}(\mathbb{I})=\mathbb{I}$. Obviously trace preserving
maps leave the set $\mathcal C_{1,1}$ of density matrices invariant.
If $\phi$ in $\mathcal J_{1}$ is completely positive and trace invariant, then by
(\ref{bound02}) $||\phi||_{1}=1$.

All completely positive maps on $\mathcal J_{1}$ have the Kraus
representation \cite{KK}, a consequence of a
theorem of Stinespring \cite{St} (see also \cite{Choi} for a proof in
the finite dimensional case): Given such a $\phi$ there is an at most
denumerable set of elements $\underline{\alpha}=\{\alpha_{i}\}_{i\in\mathbb{N}}$
in $\mathcal B(\mathcal H)$ satisfying
\begin{equation}
\label{bound1}
\sum_{i\in K}\alpha_{i}^{\star}\alpha_{i}\le ||\phi||_1\mathbb{I}
\end{equation}
for any finite subset $K\subset \mathbb{N}$
such that $\phi=\phi_{\underline{\alpha}}$, again with
\begin{equation}
\label{def1}
\phi_{\underline{\alpha}}(A)=\sum_{i\in\mathbb{N}}\alpha_{i}A\alpha_{i}^{\star},
\end{equation}
which now may be an infinite sum. This representation is not unique.
Conversely each such $\phi_{\underline{\alpha}}$ is completely
positive. More precisely $\phi_{\underline{\alpha}}$ is defined as
follows. Let
$\underline{\alpha}_{N}=(\alpha_{1},.....,\alpha_{N})$. Then
the $\phi_{\underline{\alpha}_{N}}$ form a Cauchy sequence with respect
to the norm $||\;||_{1}$ and $\phi_{\underline{\alpha}}$ is defined as the
limit (see e.g. \cite{KK}).
If $\sum_{i}\alpha_{i}^{\star}\alpha_{i}=\mathbb{I}$, such that $||\phi||_{1}=1$, then
$\phi_{\underline{\alpha}}$ is trace preserving in $\mathcal J_{1}$.
In the finite dimensional case $(\dim\mathcal H<\infty)$ the representation
may always be chosen such that the index $i$ runs through a finite set of
the order at most $(\dim\mathcal H)^{2}$.

An important and interesting feature of completely positive maps in
the context of quantum physics is that they not
necessarily map pure states, i.e. one-dimensional orthogonal
projections, into pure states.

Again we have the following representation for the
adjoint of $\phi_{\underline{\alpha}}$
\begin{equation}
\phi_{\underline{\alpha}}^{\star}(A)
=\sum_{i\in\mathbb{N}}\alpha^{\star}_{i}A\alpha_{i},
\end{equation}
i.e. $\phi^{\star}_{\underline{\alpha}}=\phi_{\underline{\alpha}^{\star}}$.
So $\phi_{\underline{\alpha}}$ is trace preserving if and only if 
$\phi_{\underline{\alpha}}^{\star}(\mathbb{I})=\mathbb{I}$ and then 
$||\phi_{\underline{\alpha}}||_{1}=1$.
Observe that to any completely positive $\phi$ with $||\phi||_{1}\le 1$
or equivalently $\phi^{\star}(\mathbb{I})\le \mathbb{I}$ ( this condition is again 
natural in the context of quantum physics,
see e.g. \cite{KK}) we may in a natural way associate a
trace preserving, completely positive map
$\hat{\phi}$ given as
$\hat{\phi}(A)=\phi(A)+\tilde{\alpha}A\tilde{\alpha}$ where
$\tilde{\alpha}=(\mathbb{I}-\phi^{\star}(\mathbb{I}))^{1/2}\ge 0$.

An example to which we shall return below is when the $\alpha_{i}$'s are
orthogonal projection operators which are pairwise orthogonal, i.e. satisfy
\begin{equation}
\label{ex1}
\alpha_{i}\alpha_{j}=\delta_{ij}\alpha_{i}=\delta_{ij}\alpha_{i}^{\star}.
\end{equation}
Then $\phi_{\underline{\alpha}}^{\star}(\mathbb{I})=\mathbb{I}$ if and
only if the $\alpha_{i}$ provide a decomposition of unity, i.e. if
$\sum_{i\in\mathbb{N}}\alpha_{i}=\mathbb{I}$ holds. As already remarked 
$\phi_{\underline{\alpha}}$ is
then also trace preserving. More generally if the $\alpha_{i}$ are all
selfadjoint with $\sum_{i\in\mathbb{N}}\alpha_{i}^{2}=\mathbb{I}$, then
$\phi_{\underline{\alpha}}$ is trace preserving.

We remark that there are some situations (see
e.g. \cite{KS} and which actually was the motivation for the present
discussion) where one
starts with completely positive maps of the following form. Let
$(\Omega,\mu)$ be a measure space, i.e. $\mu$ is countably additive.
Suppose we are given a measurable map
$\omega\rightarrow \alpha(\omega)$
from $\Omega$ into $\mathcal B(\mathcal H)$ such that
\begin{equation}
\int ||\alpha(\omega)||^{2}d\mu(\omega)< \infty.
\end{equation}
We then set
\begin{equation}
\label{def2}
\phi(A)=\int\alpha(\omega)A\alpha(\omega)^{\star}d\mu(\omega),
\end{equation}
such that its adjoint takes the form
\begin{equation}
\phi^{\star}(A)=\int\alpha(\omega)^{\star}A\alpha(\omega)d\mu(\omega),
\end{equation}
The condition corresponding to (\ref{bound1}) is given as
\begin{equation}
\label{bound3}
\phi^{\star}(\mathbb{I})=\int\alpha(\omega)^{\star}\alpha(\omega)d\mu(\omega)\le c\,\mathbb{I}
\end{equation}
for some $0<c<\infty$.
To cast $\phi$ in the form (\ref{def1}) such that (\ref{bound3})
gives (\ref{bound1}), let $\chi_{i}$ be any
orthonormal basis in the separable Hilbert space $L^{2}(\Omega, d\mu)$ and set
\begin{equation*}
\alpha_{i}=\int\overline{\chi_{i}}(\omega)\alpha(\omega)d\mu(\omega).
\end{equation*}
As an example one may choose $\Omega$ to be a compact group, $\mu$ its Haar
measure and $\omega\rightarrow\alpha(\omega)$ a unitary
representation. Then the image of $\phi$ consists of all trace class
operators which commute with all $\alpha(\omega)$. Also $\phi$ is
trace preserving and an idempotent, i.e. $\phi^{2}=\phi$. Moreover
$\phi$ is a conditional
expectation when $\dim\mathcal H<\infty$,
i.e. $\phi(A\phi(B))=\phi(A)\phi(B)=\phi(\phi(A)B)$ holds for each
$A,B$.

\section{The Perron-Frobenius Theorem}
Before we turn to a discussion of the Perron-Frobenius theorem we
make some general remarks on $\sigma_{p}(\phi)$ for the map
$\phi$ in $\mathcal J_{p}$, in particular when
$\phi$ is positive. Here $\sigma_{p}(\phi)\subseteq \mathbb{C}$ is the
spectrum of $\phi$, i.e. the set of all $\lambda$ for which
$(\lambda-\phi)$ does not have a bounded inverse in $\mathcal J_{p}$. Let
$r_{p}(\phi)$ be the
spectral radius of any bounded linear map $\phi$
in $\mathcal J_{p}$ , i.e.
\begin{equation}
\label{specrad}
r_{p}(\phi)=\lim_{n\rightarrow\infty}||\phi^{n}||_{p}^{1/n}\le ||\phi||_{p}.
\end{equation}
By a celebrated general result of Gelfand \cite{Gelfand}, one has
$\sup |\sigma_{p}(\phi)|=r_{p}(\phi)$. If the map $\phi$ in
$\mathcal J_{1}$ is trace
preserving and completely positive then
$r_{1}(\phi)=||\phi||_{1}=1$. Indeed, $\phi^{n}$ is then also completely
positive and trace preserving such that by (\ref{bound02})
$||\phi^{n}||_{1}=1$ holds for all $n$.

Lemma 2.1 allows us to draw the following conclusions on
$\sigma_{p}(\phi)$ for
positive maps $\phi$ in $\mathcal J_{p}$. First we observe that in general
$\sigma_{p}(\phi)$ is not contained in the real axis. In fact, in the
finite dimensional case $\sigma_{p}(\phi)\subseteq \mathbb{R}$ if and only if
$\phi^{\star}=\phi$ and in the infinite dimensional case the same
statement is valid when $p=2$. Let $\rho_{p}(\phi)$ be the resolvent set of the
map $\phi$ in $\mathcal J_{p}$,
i.e. the complement in $\mathbb{C}$ of $\sigma_{p}(\phi)$. If $\lambda\in
\rho_{p}(\phi)$, then to each $A\in\mathcal J_{p}$ there is a unique
$A^{\prime}\in\mathcal J_{p}$, such that $(\lambda-\phi)(A^{\prime})=A$. Taking the
adjoint and using (\ref{star3}) gives
$(\overline{\lambda}-\phi)(A^{\prime^{\star}})=A^{\star}$. Since
$A$ can be chosen arbitrary also $A^{\star}$ can be chosen arbitrary
giving a unique $A^{\star^{\prime}}$, this shows that both sets
$\rho_{p}(\phi)$
and $\sigma_{p}(\phi)$ lie symmetric with respect to the real axis.
Let $\lambda\in \sigma(\phi)$ be a real eigenvalue with eigenvector
$A$: $\phi(A)=\lambda\,A$. So by (\ref{star3}) $A^{\star}$ is
also an eigenvector with the same eigenvalue. With the decomposition
$A=A_{1}+iA_{2}$ we see that both self-adjoint operators $A_{1}$ and
$A_{2}$ are eigenvectors, so any eigenspace for a real eigenvalue is
spanned by self-adjoint elements.
If in addition the map $\phi$ in $\mathcal J_{1}$ is trace preserving and
$\lambda\neq 1$ an
eigenvalue and $A$ a corresponding eigenvector, then from
$\tr(A)=\tr(\phi(A))=\lambda\tr(A)$ we deduce $\tr(A)=0$.

In the finite dimensional case $\dim\mathcal H\,<\,\infty$ all spaces $\mathcal J_{p}$
are equal and of course $\sigma_{p}(\phi)$ is independent of $p$.
Since all norms $||\;||_{p}$ are equivalent, by its definition
$r_{p}(\phi)$ is also independent of $p$, as it should be.

To formulate the Perron-Frobenius theorem in the present context, we make the
following definition, which is the just an adaption of the usual
definition. Since any positive map
$\phi$ in $\mathcal J_{p}$ is bounded $\exp t\phi$ is a well defined positive
map in $\mathcal J_{p}$ for all $t>0$. Also its inverse $\exp -t\phi$ is a
well defined bounded map in $\mathcal J_{p}$, not necessarily positive .

\begin{defi}
A positive map $\phi$ in $\mathcal J_{p}\;(1\le p\le\infty)$ is positivity
improving if $\phi(A)\,>\,0$ for any $A\ge 0,\, A\neq 0$. $\phi$ is
ergodic if for any $A\ge 0, A\neq 0$ there is $t_{A}>0$ with
$\,(\exp t_{A}\phi)(A)\,>\,0$.
\end{defi}
 If $\phi$ is ergodic then $(\exp t\phi)(A)\;\ge (\exp t_{A}\phi)(A) >0$
for all $t\ge t_{A}$. Obviously $\phi$ is ergodic if it is
positivity improving.

A simple necessary criterion for ergodicity is given by
\begin{lem} If $\phi$ is ergodic then $Ker \phi|_{C_{p}}=0$.
\end{lem}
\begin{proof}Assume that there is $0\le A\neq 0$ with $\phi(A)=0$,
such that $(\exp t\phi)(A)=A$ for all $t$. If $A$ is not positive
definite, then we have a contradiction, so let $A>0$. We claim there is
 $0\le A^{\prime}\neq 0$ which is not positive definite with
 $A^{\prime}\le A$. In fact, we may take $A^{\prime}$ to be a
one-dimensional orthogonal projection onto an eigenvector of $A$ times
the corresponding eigenvalue, which we choose not to be zero.
This gives $0\le \phi(A^{\prime})\le \phi(A)=0$ and we are back to the
first situation.
\end{proof}

Some remarks concerning the finite dimensional case are in order.

First we will show that when $\dim\mathcal H<\infty$ the present definition
of ergodicity of $\phi$
is equivalent to the condition $(1+\phi)^{\dim\mathcal H -1}(A)>0$ for any
$0\le A\neq 0$, a criterion used in
\cite{EH}. Obviously this last condition implies ergodicity since
$\exp ta \ge 1/(n!)^{2}\min(1,t)(1+a)^{n}$ for all $n$ and all $t,a \ge
0$. As for the converse assume there is $0\le A\neq 0$ such that
$(1+\phi)^{\dim\mathcal H -1}(A)$ is not positive definite. So there is
an orthogonal projection $P\neq 0,\mathbb{I}$ and $\lambda>0$ such that
$\phi(P)\le\lambda P$ (see \cite{EH}). This gives
$(\exp t\phi)(P)\le (\exp t\lambda)P$ for all $t>0$, so $\phi$ is not
ergodic.

We have the following necessary and simple criterion for $\phi$ to be ergodic.
\begin{lem}
If $\dim\mathcal H\,<\infty$ then $\phi(\mathbb{I})>0$ for ergodic $\phi$.
\end{lem}
The converse is not true as may be seen by taking $\phi$ to be given
by (\ref{def1}), where the $\alpha_{i}$'s are taken to be a nontrivial
decomposition of unity. Then $\phi(\mathbb{I})=\mathbb{I}$ whereas
$\phi(\alpha_{i})=\alpha_{i}$ for all
$i$, such that $\phi$ is not ergodic.

\begin{proof}Assume that $\phi(\mathbb{I})\ge 0$
holds such that there is $\varphi\in\mathcal H$ with
$\langle\varphi,\phi(\mathbb{I})\varphi\rangle=0$. 
Then $\langle\varphi,\phi(A)\varphi\rangle=0$ holds
for all selfadjoint $A$ due to the bound $-||A||\mathbb{I}\le A\le ||A||\mathbb{I}$
and therefore
$\langle\varphi,\phi^{n}(A)\varphi\rangle=0$ for all $n\ge 1$ and all
$A=A^{\star}$. This in turn implies $\langle\varphi,(\exp
t\phi)(\phi(A))\varphi\rangle=0$ for all $t>0$ and all $A=A^{\star}$. Since
$\phi\neq 0$ by definition, there is $A\ge 0$ with $0\le \phi(A)\neq
0$ (see the decomposition (\ref{decomp})), so $\phi$ is not ergodic.
\end{proof}

Returning to the general case it is clear that $\phi_{\alpha}$ given by
(\ref{def1})
is not ergodic if the
$\alpha_{i}$'s satisfy
condition (\ref{ex1}). Also in the context $p=1$ and with a
continuous, one
parameter family of unitary operators $U(t)$ defining a quantum
mechanics on $\mathcal H$ in terms of its infinitesimal generator $H$, then
$\phi_{U(t)}$ (which describes the time evolution in the Heisenberg
picture) for fixed $t$ is not ergodic. If the completely
positive map $\phi$ in $\mathcal J_{p}$ is given by (\ref{def2}) in
terms of a unitary representation of a compact group, which contains
at least one irreducible representation with finite multiplicity, then by
Schur's Lemma $\phi$ is
easily seen to be ergodic ( and even positivity improving) if and only
if the representation is
irreducible. Then also $\mathcal H$ is of course finite dimensional and the
map $\phi$ is a conditional expectation.
The following two lemmas are almost obvious.
\begin{lem}
A completely positive map $\phi$ in $\mathcal J_{1}$ represented as
$\phi_{\underline{\alpha}}$ is positivity improving
if and only for any
$0\neq\varphi\in\mathcal H$ the closed linear hull of the set of vectors
$\{\alpha_{i}^{\star}\varphi\}$ is all of $\mathcal H$.
\end{lem}
\begin{proof}It suffices to consider $\phi(P)$, where $P$ is any 
one-dimensional
orthogonal projection (compare the proof of Lemma 3.1).
Let $0\neq \varphi_{0}\in P\mathcal H $ be a unit vector, so we have
to consider
\begin{equation}
\label{pocon}
\langle\varphi,\phi(P)\varphi\rangle=
\sum_{i\in\mathbb{N}}|\langle\varphi,\alpha_{i}\varphi_{0}\rangle|^{2}
\end{equation}
and the claim follows.
\end{proof}
This proof shows that in the infinite dimensional case no
$\phi_{\underline{\alpha}}$ with only finitely many non-vanishing
$\alpha_{i}$'s can be positivity improving. In fact, for given
$P$ choose $0\neq \varphi$ to be orthogonal to all
$\alpha_{i}\varphi_{0}$.
Then (\ref{pocon}) vanishes and therefore $\phi(P)$ is not positive definite.

Let $\mathcal A(\underline{\alpha})$ be the algebra (not
$^{\star}$-algebra) generated by the $\alpha_{i}^{\star}$'s.
\begin{lem}
A completely positive map $\phi$ in $\mathcal J_{1}$ represented as
$\phi_{\underline{\alpha}}$ is ergodic if and only if every
non-zero vector $\varphi$ in $\mathcal H$ is cyclic for
$\mathcal A(\underline{\alpha})$, i.e. the strong closure of the
linear space $\mathcal A
(\underline{\alpha})\varphi$ is all of $\mathcal H$.
\end{lem}
These two  results apply to all Kraus representations
$\phi_{\underline{\alpha}}$ of $\phi$.

\begin{proof}We use the same notation as in the proof of the previous lemma.
In addition write
$\alpha_{\underline{i}}=\alpha_{i_{1}}\alpha_{i{2}}..
\alpha_{i_{n}}$ for $\underline{i}=(i_{1},i_{2},..i_{n})$ and set
$|\underline{i}|=n$.
Then
\begin{equation*}
\langle\varphi,(\exp t\phi)(P)\varphi\rangle=\sum_{n=0}^{\infty}\frac{t^{n}}{n!}
\sum_{\underline{i}: |\underline{i}|=n}|
\langle\varphi,\alpha_{\underline{i}}\varphi_{0}\rangle|^{2}
\end{equation*}
and again the claim follows.
\end{proof}
The next theorem is part of the Perron-Frobenius theorem in the
present context. It is part because it assumes that $r_{p}(\phi)$ is
an eigenvalue, a standard assumption one has to make in the infinite
dimensional context if no other information is available.

\begin{theo}
Let the positive map $\phi$ in $\mathcal J_{p}\;(1\le p\le\infty$) be
ergodic and assume that
$r_{p}(\phi)=||\phi||_{p}$. If $r_{p}(\phi)$
is an eigenvalue of $\phi$, then this eigenvalue is simple and the
eigenvector $A$ may be chosen to be positive definite.
\end{theo}
\begin{re}
According to usual terminology such a vector is called a
Perron-
Frobenius vector. When $\dim{\mathcal H}<\infty$ then $r_{p}(\phi)$
is independent of $p$ and
is always an eigenvalue. This case has been covered by Evans
and Hoegh-Krohn \cite{EH} with an application to quantum Monte-Carlo
processes and by Groh \cite{Groh}. At present we do not know whether in the infinite
dimensional case the condition $r_{p}(\phi)=||\phi||_{p}$ can be dropped.
\end{re}
For the proof of the theorem we need the following simple
\begin{lem}
For any $B>0$ and $0\neq B^{\prime}\ge 0$ in $\mathcal J_{p},\,1\le p\le\infty$
the following strict inequality
\begin{equation}
\label{pos1}
||B-B^{\prime}||_{p}<||B+B^{\prime}||_{p}
\end{equation}
holds.
\end{lem}
The condition $B>0$ cannot be weakened to $B\ge 0$ as the
choice $B=P\neq 0$ and $B^{\prime}=Q\neq 0$ for two finite dimensional
orthogonal projections with $PQ=0$ shows.
\begin{proof}The estimate (\ref{pos1}) for the case $p=\infty$ is trivial, so let
$p<\infty$. Since $B-B^{\prime}\in \mathcal J_{p}$ is self-adjoint, there is a
complete set orthonormal eigenvectors $\varphi_{n}$ with
real eigenvalues $\lambda_{n}$ of $B-B^{\prime}$. With
$\lambda_{n}=\langle\varphi_{n},(B-B^{\prime})\varphi_{n}\rangle$ we obtain
\begin{equation}
\label{pos3}
||B-B^{\prime}||_{p}^{p}=\sum_{n}|\lambda_{n}|^{p}
=\sum_{n}|\langle\varphi_{n},B\varphi_{n}\rangle
-\langle\varphi_{n},B^{\prime}\varphi_{n}\rangle|^{p}.
\end{equation}
Observe that
\begin{eqnarray}
\label{pos4}
|\langle\varphi_{n},B\varphi_{n}\rangle-
\langle\varphi_{n},B^{\prime}\varphi_{n}\rangle|^{p}&\le&
\max(\langle\varphi_{n},B\varphi_{n}\rangle^{p},
\langle\varphi_{n},B^{\prime}\varphi_{n}\rangle^{p})
\nonumber\\
&\le&(\langle\varphi_{n},B\varphi_{n}\rangle+
\langle\varphi_{n},B^{\prime}\varphi_{n}\rangle)^{p}
\nonumber\\
&=&\langle\varphi_{n},(B+B^{\prime})\varphi_{n}\rangle^{p}\nonumber\\
&\le & \langle\varphi_{n},(B+B^{\prime})^{p}\varphi_{n}\rangle
\end{eqnarray}
holds, since $B>0$ and $B^{\prime}\ge 0$.
In the last estimate in (\ref{pos4}) we used the estimate (\ref{pos40}).

Furthermore
$\langle\varphi_{n},B\varphi_{n}\rangle\,>0$ holds for all $n$. Also
$\langle\varphi_{n},B^{\prime}\varphi_{n}\rangle\,>0$ for at least one
$n$. Otherwise we would also have
$\langle\varphi_{n},B^{\prime}\varphi_{n^{\prime}}\rangle=0$ for 
all $n,n^{\prime}$ by
Schwarz inequality and hence $B^{\prime}=0$, contradicting the assumption.
Thus the first inequality in (\ref{pos4}) is actually strict for at
least one $n$. Inserting (\ref{pos4}) into (\ref{pos3})
and using this last observation proves (\ref{pos1}).
\end{proof}
$Proof\; of\; the\; Theorem$:
Let $A\neq 0$ be any eigenvector for $\phi$ with eigenvalue
$r_{p}(\phi)$. By an observation made above, we may assume that $A$ is
self-adjoint. Observe the following estimate, valid for any $t>0$.
\begin{eqnarray}
\label{pos5}
\exp tr_{p}(\phi)||A||_{p}&=&||\exp tr_{p}(\phi)A||_{p}
=||(\exp t\phi)(A)||_{p}\nonumber\\
&=&||\,|(\exp t\phi)(A)|\,||_{p}\le||(\exp t\phi)(|A|)||_{p}\nonumber\\
&\le& \exp t||\phi||_{p}||\,|A|\,||_{p}
=\exp t||\phi||_{p}||A||_{p}.
\end{eqnarray}
Here we have used the estimate (\ref{mon}) in combination with
(\ref{mon1}) for the map $\exp t\phi$.

Since by assumption $r_{p}(\phi)=||\phi||_{p}$, all inequalities in
(\ref{pos5}) actually have to be equalities. In particular for all $t>0$
\begin{eqnarray}
\label{pos6}
||(\exp t\phi)(A)||_{p}&=&||(\exp t\phi)(A_{+})-(\exp t\phi)(A_{-})||_{p}
\nonumber\\
&=&||(\exp t\phi)(|A|)||_{p}=||(\exp t\phi)(A_{+})+(\exp t\phi)(A_{-})||_{p}.
\end{eqnarray}
If $A_{+}=0$ or $A_{-}=0$ there is nothing to prove so assume that
$A_{\pm}\neq 0$. By the ergodicity of $\phi$  we have
$(\exp t\phi)(A_{\pm})>0$ for all large $t>0$.
But then (\ref{pos6}) contradicts (\ref{pos1}). So we must have
either $A_{+}=0$ or
$A_{-}=0$ and by replacing $A$ by $-A$ if necessary we may without
restriction assume that $A=|A|=A_{+}$. Again by ergodicity we have
$\exp tr_{p}(\phi)A=(\exp t\phi)(A)>0$ for all large $t$ such that $A>0$.
The proof of non-degeneracy is now easy. Assume there are two linearly
independent eigenvectors $A_{1}>0$ and $A_{2}>0$, which we may normalize to
$||A_{1}||_{p}=||A_{2}||_{p}=1$. Then $0\neq A_{1}-A_{2}$ is also an
eigenvector and by our previous discussion we must have either
$A_{1}>A_{2}$ or $A_{2}>A_{1}$. This would imply that
$||A_{1}||_{p}>||A_{2}||_{p}$ or $||A_{2}||_{p}>||A_{1}||_{p}$
respectively, again an easy consequence of (\ref{mon}) and its proof.
This is a contradiction, so we must have $A_{1}=A_{2}$, thus concluding the
proof of the theorem.

\section{Existence of eigenvalues}

In this section we want in particular to establish conditions for
positive maps
$\phi$ in $\mathcal J_{p}$, which are sufficient to show that $r_{p}(\phi)$ is an
eigenvalue. We start with a preparation. For any $A\in\mathcal B(\mathcal H)$ we
write $A=U_{A}|A|$ with unique $U_{A}$ for its polar decomposition.
More precisely $U_{A}$ is isometric on $\overline{\Ran |A|}=(Ker
A)^{\perp}$, i.e. $U_{A}^{\star}U_{A}\varphi=\varphi$ for
$\varphi\in\overline{\Ran |A|}=(Ker A)^{\perp}$, and $U_{A}\varphi=0$
for $\varphi\in (\Ran |A|)^{\perp}=Ker A$.

The next two lemmas replace the Kadison-Schwarz inequality
(see e.g. the second reference in \cite{Kadison} and
\cite{Stoermer,Choi1,LR} for other early references) used in
the context of $C^{\star}$-algebras in \cite{EH,Groh}  and in the context of
von Neumann algebras in \cite{AH}. Since the spaces $\mathcal J_{p},\,
p<\infty$ are not algebras these two lemmas will provide the
appropriate substitute.

\begin{lem} Let the map $\phi$ in $\mathcal J_{p},\;1\le p\le\infty$ be 2-positive.
Then for all $\varphi,\varphi^{\prime}\in\mathcal H$ and all $A\in\mathcal J_{p}$
\begin{equation}
\label{bound5a}
|\langle\varphi^{\prime},\phi(A)\varphi\rangle|^{2}\le
\langle\varphi,\phi(|A|)\varphi\rangle
\langle\varphi^{\prime},\phi\circ\phi_{U_{A}}(|A|)\varphi^{\prime}\rangle.
\end{equation}

Also if not all terms in (\ref{bound5a}) are vanishing, then there is
equality if and only if both relations
\begin{equation}
\label{mat0}
\left(\begin{array}{cc}\phi(|A|)&\phi(A^{\star})\\
                       \phi(A)&\phi\circ\phi_{U_{A}}(|A|)
\end{array} \right)\left(\begin{array}{cc}
  \langle\varphi^{\prime},\phi\circ\phi_{U_{A}}(|A|)\varphi^{\prime}\rangle&\varphi\\
-\langle\varphi^{\prime},\phi(A)\varphi\rangle&\varphi^{\prime}
\end{array}\right)
=0
\end{equation}
and
\begin{equation}
\label{mat}
\left(\begin{array}{cc}\phi(|A|)&\phi(A^{\star})\\
                       \phi(A)&\phi\circ\phi_{U_{A}}(|A|)
\end{array} \right)\left(\begin{array}{cc}
  &\langle\varphi,\phi(A)^{\star}\varphi^{\prime}\rangle\varphi\\
-&\langle\varphi,\phi(|A|)\varphi\rangle\varphi^{\prime}
\end{array}\right)
=0
\end{equation}
hold.
\end{lem}
For $A=A^{\star}$  estimate (\ref{bound5a}) follows from
(\ref{mon1}), since
then $U_{A}^{\star}$ commutes with $A$ such that
$U_{A}|A|U_{A}^{\star}=AU_{A}^{\star}=
U_{A}{^\star}A=U_{A}^{\star}U_{A}|A|=|A|$.

\begin{proof}Here and in the proof of the next lemma we will mimic and extend
Exercise 3.4 in \cite{Paulsen}, page 39 in
the present context.
Set $A_{1}=|A|^{1/2},\,A_{2}=|A|^{1/2}U_{A}^{\star}$
such that $A_{1}^{2}=|A|,\,A_{2}^{\star}A_{1}=A, A_{2}^{\star}A_{2}=
U_{A}|A|U_{A}^{\star}=\phi_{U_{A}}(|A|)$ are all in $\mathcal J_{p}$.
Consider
\begin{equation}
\label{mat1}
\left(\begin{array}{cc}
                       A_{1}&A_{2}\\
                       0&0
\end{array}\right)^{\star}
\left(\begin{array}{cc}
                       A_{1}&A_{2}\\
                       0&0
\end{array} \right)=\left(\begin{array}{cc}|A|&A^{\star}\\
                       A&\phi_{U_{A}}(|A|)
\end{array} \right)\ge 0
\end{equation}
in $\mathcal H\oplus\mathcal H\cong \mathcal H\otimes\mathbb{C}^{2}$, which defines an element in
$\mathcal C_{p}(\mathcal H\oplus\mathcal H)$. By assumption
\begin{equation}
\label{mat2}
\phi_{2}\left(\begin{array}{cc}
                      |A|&A^{\star}\\
                       A&\phi_{U_{A}}(|A|)
\end{array} \right)=\left(\begin{array}{cc}\phi(|A|)&\phi(A^{\star})\\
                       \phi(A)&\phi\circ\phi_{U_{A}}(|A|)
\end{array} \right)\ge 0.
\end{equation}
Consider the following linear transformation $T$ on $\mathbb{C}^{2}$
(equipped with the natural scalar product) given as the $2\times 2$ matrix
\begin{equation}
T=\left(\begin{array}{cc}\langle\varphi,\phi(|A|)\varphi\rangle&
\langle\varphi,\phi(A)^{\star}\varphi^{\prime}\rangle\\
\langle\varphi^{\prime},\phi(A)\varphi\rangle&
\langle\varphi^{\prime},\phi\circ\phi_{U_{A}}(|A|)\varphi^{\prime}\rangle\end{array}\right)
\end{equation}
such that for all $a_1,a_2\in \mathbb{C}$
\begin{equation}
\label{tphi}
\langle\left(\begin{array}{c} a_{1}\\
                        a_{2}\end{array}\right),\; T
                        \left(\begin{array}{c}
                        a_{1}\\
                        a_{2}\end{array}\right)\rangle=
\langle\left(\begin{array}{c} a_{1}\;\varphi\\
                        a_{2}\;\varphi^{\prime}\end{array}\right)
,\;\left(\begin{array}{cc}\phi(|A|)&\phi(A^{\star})\\
                       \phi(A)&\phi\circ\phi_{U_{A}}(|A|)
\end{array} \right)
             \left(\begin{array}{c} a_{1}\;\varphi\\
                        a_{2}\;\varphi^{\prime}\end{array}\right)\rangle\;\ge 0
\end{equation}
with the obvious notation for the scalar product in $\mathcal H\oplus\mathcal H$.
A linear transformation in $\mathbb{C}^{2}$ is $\ge 0$ if and only if its
trace and its determinant are both $\ge 0$. Hence the bound (\ref{bound5a})
follows. This discussion also easily gives the last claim in Lemma
4.1. Indeed, if $T\neq 0$, its determinant is equal to zero if
and only if there is a non-zero eigenvector of $T$ with eigenvalue zero and
this is the case if and only if equality in (\ref{bound5a}) holds and
then such an eigenvector of $T$ is given either as
\begin{equation*}
\left(\begin{array}{c}
  \langle\varphi^{\prime},\phi\circ\phi_{U_{A}}(|A|)\varphi^{\prime}\rangle\\
                        -\langle\varphi^{\prime},\phi(A)\varphi\rangle\end{array}\right)
\end{equation*}
or as
\begin{equation*}
\left(\begin{array}{c}
  \langle\varphi,\phi(A)^{\star}\varphi^{\prime}\rangle\\
 -\langle\varphi,\phi(|A|)\varphi\rangle \end{array}\right).
\end{equation*}
More precisely, if all matrix elements of $T$ are non-vanishing then these
two eigenvectors have non-zero entries and are proportional. If
$T\neq 0$ with $\det T=0$ and if one diagonal element is vanishing,
such that also the
off diagonal elements are vanishing, then exactly one of these vectors
is the null vector.
In view of (\ref{tphi}) this concludes the proof of Lemma 4.1.
\end{proof}
\begin{lem}
Let the map $\phi$ in $\mathcal J_{p},\;1\le p\le\infty$ be 2-positive.
Then the bound
\begin{equation}
\label{bound5}
||\phi(A)||_{p}\le||\phi(|A|)||_{p}^{1/2}\;\;
||\phi\circ\phi_{U_{A}}(|A|)||_{p}^{1/2}
\end{equation}
holds for all $A\in\mathcal J_{p}$. In particular $A\in Ker \phi$ if
$|A|\in Ker\phi$ or $|A|\in Ker \phi\circ\phi_{U_{A}}$.
\end{lem}

\begin{proof}Choosing $\varphi^{\prime}=U_{\phi(A)}\varphi$ in (\ref{bound5a}) we have
the estimate
\begin{equation}
\label{bound6}
\langle\varphi,|\phi(A)|\varphi\rangle\le \langle\varphi,\phi(|A|)\varphi\rangle^{1/2}
\langle U_{\phi(A)}\varphi,\phi\circ\phi_{U_{A}}(|A|)U_{\phi(A)}\varphi\rangle^{1/2}
\end{equation}
for all $\varphi\in\mathcal H$. Thus (\ref{bound5}) follows trivially from
(\ref{bound6}) in case $p=\infty$. So let
$1\le p\le\infty$.
Take $\varphi_{n}$ to be an orthonormal basis in $\mathcal H$ diagonalizing
$|\phi(A)|$ with eigenvalues $\mu_{n}\ge 0$.
Then we obtain
\begin{eqnarray}
\label{bound7}
||\phi(A)||_{p}^{p}&=&\sum_{n}\mu_{n}^{p}=
  \sum_{n}\langle\varphi_{n},|\phi(A)|\varphi_{n}\rangle^{p}\nonumber\\
                   &\le
                   &\sum_{n}\langle\varphi_{n},\phi(|A|)\varphi_{n}\rangle^{p/2}
\langle U_{\phi(A)}\varphi_{n},
\phi\circ\phi_{U_{A}}(|A|)U_{\phi(A)}\varphi_{n}\rangle^{p/2}\nonumber\\
&\le&\left(\sum_{n}\langle\varphi_{n},\phi(|A|)^{p}\varphi_{n}\rangle\right)^{1/2}\cdot\nonumber\\
&&\left(\sum_{n}\langle U_{\phi(A)}\varphi_{n},
(\phi\circ\phi_{U_{A}}(|A|))^{p}U_{\phi(A)}\varphi_{n}\rangle\right)^{1/2}\nonumber\\
&\le&||\phi(|A|)||_{p}^{p/2}\:\:||\phi\circ\phi_{U_{A}}(|A|)||_{p}^{p/2}.
\end{eqnarray}
Here we have used (\ref{bound6}) for the first inequality. The
second inequality is a consequence of Schwarz inequality and
(\ref{pos40}). In what follows this will become important, since we
will encounter the situation when all inequalities are actually
equalities. Also we have
used the fact that with $\varphi_{n}$ being an orthonormal basis also
$U_{\phi(A)}\varphi_{n}$ is a possibly incomplete set of orthonormal
vectors and which is responsible for the last inequality.
\end{proof}
$\lambda\in\mathbb{C}$ is called a peripheral eigenvalue of the positive map $\phi$ in
$\mathcal J_{p}$ if $\lambda$ and hence also $\overline{\lambda}$ is an
eigenvalue with $|\lambda|=r_{p}(\phi)$.
\begin{theo}
Let the map $\phi$ in $\mathcal J_{p},\;1\le p<\infty$ be 2-positive with
$r_{p}(\phi)=||\phi||_{p}$. Assume that $\lambda\in\mathbb{C}$ is a peripheral
eigenvalue with eigenvector $A$. Then $r_{p}(\phi)$ is also
an eigenvalue with eigenvector $|A|$. In particular, if $\phi$ is
2-positive and compact with
$r_{p}(\phi)=||\phi||_{p}$, then $r_{p}(\phi)$ is an eigenvalue.
\end{theo}
Here and in contrast to the previous discussions the case $p=\infty$
is not covered by the present discussion. Since $\mathcal J_{p=\infty}$ is a
$C^{\star}$-algebra, this case, however, is covered by the discussion
in \cite{WaEn,Groh}. Then and similarly in the context of von Neumann
algebras one can actually prove more. In fact, if the spectrum is rescaled via
$\lambda\rightarrow\lambda/r_{p}(\phi)$, then the peripheral
eigenvalues form a discrete sub-group of the unit circle group, see
\cite{AH} and the first reference in \cite{Groh}.

\begin{proof} We want to
prove that $|A|\neq 0$ is also an eigenvector with eigenvalue
$|\lambda|=r_{p}(\phi)$, i.e. $\phi(|A|)=r_{p}(\phi)|A|$. Since
$|\phi(A)|=|\lambda A|=r_{p}(\phi)|A|$ (such that also
$U_{\phi(A)}=(\lambda/|\lambda|)U_{A}$ and hence
$\phi_{U_{\phi(A)}}=\phi_{U_{A}}$ by the uniqueness of $U_{A}$)
it suffices to show that
\begin{equation}
\label{eigen}
\phi(|A|)\varphi_{n}=\mu_{n}\varphi_{n}
\end{equation}
holds for all $n$, where the $\varphi_{n}$ are as in the previous lemma.
Also by the previous lemma we have
\begin{eqnarray}
r_{p}(\phi)||A||_{p}&=&||\lambda
A||_{p}=||\phi(A)||_{p}=||\,|\phi(A)|\,||_{p}\nonumber\\
 &\le&||\phi(|A|)||_{p}^{1/2}
 ||\phi\circ\phi_{U_{A}}(|A|)||_{p}^{1/2}\nonumber\\
   &\le&||\phi||_{p}\;\;||A||_{p}.
\end{eqnarray}
Here we have used the fact that $||\phi_{U_{A}}||_{p}\le 1$.
By the assumption $r_{p}(\phi)=||\phi||_{p}$ we must have equality. In
particular this implies that
\begin{equation}
\label{eq1}
||\,|\phi(A)|\,||_{p}
 =||\phi(|A|)||_{p}^{1/2}
 ||\phi\circ\phi_{U_{A}}(|A|)||_{p}^{1/2}.
\end{equation}
Inspection of the proof of Lemma 4.2 shows that we must
have equality
in (\ref{bound6}) when $\varphi=\varphi_{n}$ for all $n$. The
relations (\ref{mat0}) and (\ref{mat}) in Lemma
4.1 then imply that we must have
\begin{equation}
\label{eqphin1}
\langle\varphi_{n}^{\prime},\phi\circ\phi_{U_{A}}
(|A|)\varphi_{n}^{\prime}\rangle
\phi(|A|)\varphi_{n}=
\langle\varphi_{n}^{\prime},\phi(A)\varphi_{n}\rangle
\phi(A)^{\star}\varphi_{n}^{\prime}
\end{equation}
and
\begin{equation}
\label{eqphin2}
\langle\varphi_{n},\phi(|A|)\varphi_{n}\rangle
\phi\circ\phi_{U_{A}}(|A|)\varphi_{n}^{\prime}=
\langle\varphi_{n},\phi(A)^{\star}\varphi_{n}^{\prime}\rangle\phi(A)\varphi_{n}
\end{equation}
for all $n$ with $\varphi_{n}^{\prime}=U_{\phi(A)}\varphi_{n}$.
Assume $n$ to be such that $\mu_{n}>0$. Since $\varphi_{n}\in \Ran |\phi(A)|$
we have
$U_{\phi(A)}^{\star}\varphi_{n}^{\prime}=\varphi_{n}$ and
$\phi(A)^{\star}\varphi_{n}^{\prime}=|\phi(A)|\varphi_{n}=\mu_{n}\varphi_{n}$
for all such $n$. Similarly
$\phi(A)\varphi_{n}=U_{\phi(A)}|\phi(A)|\varphi_{n}
=\mu_{n}\varphi_{n}^{\prime}$.
Hence we may rewrite (\ref{eqphin1}) and (\ref{eqphin2}) as
\begin{equation}
\label{ev1}
\langle\varphi_{n}^{\prime},\phi\circ\phi_{U_{A}}
(|A|)\varphi_{n}^{\prime}\rangle
\phi(|A|)\varphi_{n}=\mu_{n}^{2}\varphi_{n}
\end{equation}
and
\begin{equation}
\label{ev2}
\langle\varphi_{n},\phi(|A|)\varphi_{n}\rangle
\phi\circ\phi_{U_{A}}(|A|)\varphi_{n}^{\prime}=\mu_{n}^{2}\varphi_{n}^{\prime}
\end{equation}
whenever $\mu_{n}>0$.

Furthermore we recall that we made use of
Schwarz inequality in the estimate (\ref{bound7}). For this to be an equality
we claim that we actually must have
\begin{equation}
\label{ev3}
\langle\varphi_{n},\phi(|A|)\varphi_{n}\rangle=\langle\varphi_{n}^{\prime},\phi\circ\phi_{U_{A}}
(|A|)\varphi_{n}^{\prime}\rangle
\end{equation}
for all $n$. Indeed, the Schwarz inequality in this
context is an equality if and only if both sides of (\ref{ev3}) are
proportional with a common
proportionality factor for all $n$. By the equality (\ref{eq1})
this proportionality factor has to be equal to 1, thus proving
(\ref{ev3}).

If $\mu_{n}=0$ then $\varphi_{n}^{\prime}=0$ and hence by (\ref{ev3})
$\phi(A)\varphi_{n}=0$. Thus we have established (\ref{eigen}) in the
case when $\mu_{n}=0$. If $\mu_{n}>0$ then the
left hand sides of (\ref{ev1}) and (\ref{ev2}) are both
non-vanishing since then $\varphi_{n}^{\prime}\neq 0$. Therefore
$\varphi_{n}$ is an eigenvector of $\phi(|A|)$ with eigenvalue
\begin{eqnarray*}
\langle\varphi_{n},\phi(|A|)\varphi_{n}\rangle&=&
\frac{\mu_{n}^{2}}{\langle\varphi_{n}^{\prime},
\phi\circ\phi_{U_{A}}(|A|)\varphi_{n}^{\prime}\rangle}\\
&=&\frac{\mu_{n}^{2}}{\langle\varphi_{n},\phi(|A|)\varphi_{n}\rangle}.
\end{eqnarray*}
This gives (\ref{eigen}) in the case when $\mu_{n}>0$, thus
concluding the proof of the theorem, since for compact operators the spectrum
is pure point.
\end{proof}
The use of
(\ref{pos40}) in the second inequality in (\ref{bound7}) does not
give any additional mileage when $p\neq 1$
since the $\varphi_{n}$ are eigenvectors of
$\phi(|A|)$ and the $\varphi_{n}^{\prime}$ eigenvectors of
$\phi\circ\phi_{U_{A}}$ and hence (\ref{pos40}) gives an equality for
the case at hand.

The strategy of the proof was motivated by the following observation in
the classical context. With the notation introduced after (\ref{mon1})
assume that $S\underline{z}=\lambda\underline{z}$. This gives
$|\lambda|\;|\underline{z}|=|\lambda\underline{z}|=
|S\underline{z}|\le S|\underline{z}|$. Taking its usual Hilbert space
norm in $\mathbb{C}^{n}$ we obtain $|\lambda|\,||\;|\underline{z}|\;||\le
||S|\underline{z}|\,||\le ||S||\,||\,|\underline{z}|\,||$. So if
$|\lambda|=||S||$ we must have $S|\underline{z}|=||S||\;|\underline{z}|$.

The next result extends Theorem 3.1.
\begin{theo} Let the map $\phi$ in $\mathcal J_{p},\;1\le p<\infty$ be
2-positive and ergodic with $r_{p}(\phi)=||\phi||_{p}$. Assume that
$\lambda\in\mathbb{C}$ is a peripheral eigenvalue. Then
$\lambda$ is a simple eigenvalue and the corresponding eigenvector $A$
is a normal operator.
\end{theo}
\begin{proof}
Assume there are two eigenvectors $A_{1}$ and $A_{2}$ in $\mathcal J_{p}$
with eigenvalue $\lambda$, normalized to
$||A_{1}||_{p}=||\;|A_{1}|\;||_{p}=||A_{2}||_{p}=||\;|A_{2}|\;||_{p}=1$.
By Theorem 3.1, its proof and by Theorem 4.1 we have that
$0<|A_{1}|=|A_{2}|$ is an eigenvector with eigenvalue
$r_{p}(\phi)$. In particular both $U_{A_{1}}$ and $U_{A_{2}}$ are
unitaries. Since $A_{1}-\exp i\tau\;A_{2}$ is also an eigenvector for
the same eigenvalue for all $0\le\tau\le2\pi$ by the same argument we
must have
\begin{equation}
\label{a1a2}
|A_{1}-\exp i\tau\;A_{2}|=||A_{1}-\exp i\tau\;A_{2}||_{p}|A_{1}|.
\end{equation}
The function $\tau\mapsto ||A_{1}-\exp i\tau\;A_{2}||_{p}$ is
easily seen to be continuous. Hence there is $\tau_{0}$ such that
\begin{equation}
\label{a1a22}
||A_{1}-\exp i\tau_{0}\;A_{2}||_{p}
=\min_{0\le\tau\le 2\pi}||A_{1}-\exp i\tau\;A_{2}||_{p}.
\end{equation}
Take the square of (\ref{a1a2}) and use (\ref{a1a22}) to obtain
\begin{equation*}
|A_{1}-\exp i\tau_{0}\;A_{2}|^{2}\le |A_{1}-\exp i\tau\;A_{2}|^{2}.
\end{equation*}
Due to the relation $\overline{\Ran\;|A_{1}|}=\mathcal H$ this gives
\begin{equation*}
||(U_{A_{1}}-\exp i\tau_{0}\;U_{A_{2}})\varphi||^{2}
\le||(U_{A_{1}}-\exp i\tau\;U_{A_{2}})\varphi||^{2}
\end{equation*}
which written out in turn gives
\begin{equation}
\label{a1a23}
\Re\left( e^{i\tau}\langle U_{A_{1}}\varphi,U_{A_{2}}\varphi\rangle\right)
\le\Re\left( e^{i\tau_{0}}\langle U_{A_{1}}\varphi,U_{A_{2}}\varphi\rangle\right)
\end{equation}
for all $0\le\tau\le 2\pi$ and all $\varphi\in\mathcal H$. We distinguish two
possible cases. First if $A_{1}-\exp i\tau_{0}\;A_{2}=0$ and hence
$U_{A_{1}}-\exp i\tau_{0}\;U_{A_{2}}=0$ there is nothing to
prove. Otherwise $U=U_{A_{1}}^{-1}\exp i\tau_{0}\;U_{A_{2}}$ is a unitary
operator $\neq \mathbb{I}$ and hence by the spectral theorem for unitary
operators there is $\varphi\in\mathcal H$ such that
$\Re(\langle\varphi,U\varphi\rangle)<|\langle\varphi,U\varphi\rangle|$. Choose
$\tau$ such that
$\exp i(\tau-\tau_{0})\langle\varphi,U\varphi\rangle=|\langle\varphi,U\varphi\rangle|$. This
contradicts (\ref{a1a23}), thus concluding the proof of the first part of
the theorem. As for the last part $U_{A}$ is unitary, as already noted.
Relation (\ref{ev2}) shows that $\phi_{U_{A}}(|A|)$
is also an eigenvector of
$\phi$ with eigenvalue $r_{p}(\phi)$. Hence $\phi_{U_{A}}(|A|)=|A|$ by
uniqueness and normalization. So $U_{A}$ commutes with $|A|$ and therefore
$A$ is normal with $U^{\star}_{A}=U_{A^{\star}}$ and $|A^{\star}|=|A|$, where we recall that $A^{\star}$ is an eigenvector with eigenvalue
$\overline{\lambda}$.
\end{proof}

The next result is of relevance in quantum physics. We recall that for
completely positive, trace preserving maps $\phi$ in $\mathcal J_{1}$  one has
$r_{1}(\phi)=||\phi||_{1}=1$. So we immediately obtain the
following existence result for Perron-Frobenius vectors.
\begin{cor}
Let the map $\phi$ in $\mathcal J_{1}$ be completely positive, trace
preserving, compact and ergodic. Then there is a unique density matrix $\rho
>0$ which is left invariant under $\phi$.
\end{cor}
A converse of this result is given by Theorem 5.1 below.

The following result is of the type needed in the standard context of Monte
Carlo simulations in lattice models of statistical mechanics. It
states that, starting from any non-negative initial
configuration, by iterations of an ergodic up-dating procedure (heat
bath method) one
eventually arrives at the desired equilibrium configuration which
typically is given by a Gibbs distribution and which by
construction of the up-dating is a
Perron-Frobenius vector for the up-dating procedure.

\begin{cor} Let the map $\phi$ in $\mathcal J_{2}$ be 2-positive,
selfadjoint, compact and ergodic with Perron-Frobenius vector
$A>0$ normalized to $||A||_{2}=1$. Then for any $B\in\mathcal J_{2}$
\begin{equation}
\label{approach1}
s-\lim_{t\rightarrow\infty}\exp(t(\phi-||\phi||_{2}))(B)
=\langle A,B\rangle_{2}A.
\end{equation}
\end{cor}
If $0\le B\neq 0$ then the right hand side of (\ref{approach1})
is non-vanishing. It would be interesting to find an analogous
formulation when $p\neq 2$.
\begin{proof}
We use familiar arguments. Let the $A_{j}$ form an orthonormal basis
of eigenvectors for $\phi$ arranged such that $A_{1}=A$ and
$\sigma_{j+1}\le\sigma_{j}$. Here the $\sigma_{j}$'s are the
eigenvalues, i.e. $\phi(A_{j})=\sigma_{j}A_{j}$. Since
$\sigma_{1}=||\phi||_{2}$ is a simple eigenvalue, $\sigma_{1}>\sigma_{2}$.
By Plancherel's theorem we
have $B=\sum_{j}b_{j}A_{j}$ with $b_{j}=\langle A_{j},B\rangle_{2}$ and
$\sum_{j}b_{j}^{2}=||B||_{2}^{2}$. Obviously
\begin{equation*}
\exp(t(\phi-||\phi||_{2}))(B)=\sum_{j}b_{j}e^{t(\sigma_{j}-\sigma_{1})}A_{j}
\end{equation*}
holds such that
\begin{equation}
\label{approach2}
||(\exp(t(\phi-||\phi||_{2})))(B)-\langle A,B\rangle_{2}A||_{2}\le
e^{-t(\sigma_{1}-\sigma_{2})}||B||_{2}
\end{equation}
and the claim follows.
\end{proof}
By a slight modification of the proof using the spectral
representation of $\phi$ one may replace the compactness
condition by the condition that $||\phi||_{2}$ is an isolated
eigenvalue, i.e. $||\phi||_{2}$ is separated by a distance $m^{2}$
(the ``mass gap'' in physical language) from the remainder of the
spectrum of $\phi$, which
otherwise may be arbitrary. Then the estimate
(\ref{approach2}) is still valid with $\sigma_{1}-\sigma_{2}$ being
replaced by $m^{2}$.

\section{Examples}
In this section we will provide non-trivial examples when $\dim\mathcal H=\infty$.
The issue is to show that all 4 conditions in Corollary 4.1 can be
fulfilled simultaneously. In the finite dimensional case the
compactness criterion is automatically satisfied and then it is easy to
construct non-trivial examples.
In general the existence of completely positive, trace preserving,
compact maps in $\mathcal J_{1}$ is clarified by
\begin{lem}
Let $\underline{\alpha}=\{\alpha_{i}\}_{i\in\mathbb{N}}$ be such that
$\sum_{i\in\mathbb{N}}\alpha_{i}^{\star}\alpha_{i}=\mathbb{I}$ and all
$\alpha_{i}\in\mathcal B(\mathcal H)$ are compact. Then the map
$\phi_{\underline{\alpha}}$ in $\mathcal J_{1}$ is compact.
\end{lem}
\begin{proof}We recall that $\phi_{\underline{\alpha}}$ is the norm limit of
$\phi_{\underline{\alpha}_{N}}$ as $N\rightarrow\infty$. Since the
space of compact operators is closed w.r.t. the norm topology it
suffices to prove that all $\phi_{\underline{\alpha}_{N}}$ are
compact. By assumption to each $i$ there is a sequence
$\alpha_{i,k}$ of finite rank operators in $\mathcal B \mathcal H)$ such that
\begin{equation*}
 \lim_{k\rightarrow\infty}||\alpha_{i}-\alpha_{i,k}||
=\lim_{k\rightarrow\infty}||\alpha_{i}^{\star}-\alpha_{i,k}^{\star}||=0.
\end{equation*}
Since $\alpha_{i,k}^{\star}$ is of finite rank
$\phi_{\underline{\alpha}_{N,k}}$ with
$\underline{\alpha}_{N,k}=(\alpha_{1,k},...,\alpha_{N,k})$ is also of
finite rank and in particular compact. Using the a priori estimate
$||ABC||_{1}\le ||A||\,||B||_{1}\,||C||$ and a standard $\epsilon/3$
argument gives
\begin{equation*}
\lim_{k\rightarrow\infty}
||\phi_{\underline{\alpha}_{N}}-\phi_{\underline{\alpha}_{N,k}}||_{_1}=0.
\end{equation*}
\end{proof}
We are now prepared to provide non-trivial examples, which resulted
from a discussion with D. Buchholz. Let $\psi_{i},\;i\in\mathbb{N}$ be any orthonormal
basis in $\mathcal H$ and $c_{ik}>0$ any set of numbers which
satisfy the condition $\sum_{i} c_{ik}^{2}=1$ for all $k$. In the
Dirac notation we define the following family
$\underline{\alpha}=\{\alpha_{ik}\}_{1\le i,k<\infty}$ of rank 1 operators
\begin{equation*}
\alpha_{ik}=c_{ik}|\psi_{i}\rangle\langle\psi_{k}|,
\end{equation*}
such that
\begin{equation*}
\alpha_{ik}^{\star}=c_{ik}|\psi_{k}\rangle\langle\psi_{i}|,
\end{equation*}
In other words $\alpha_{ik}$ is the map
$\varphi\mapsto c_{ik}\langle\psi_{k},\varphi\rangle\psi_{i}$.
We set
\begin{equation*}
\phi_{\underline{\alpha}}(A)=\sum_{i,k}\alpha_{ik}A\alpha_{ik}^{\star}
\end{equation*}
By assumption we have
\begin{equation*}
\sum_{i,k}\alpha_{ik}^{\star}\alpha_{ik}=
\sum_{i,k}c_{ik}^{2}\;|\psi_{k}\rangle\langle\psi_{k}|=\sum_{k}|\psi_{k}\rangle\langle\psi_{k}|=\mathbb{I},
\end{equation*}
so $\phi_{\underline{\alpha}}$ is completely positive, compact and trace
preserving. We claim that it is also positivity improving by Lemma
3.3. Indeed, given $0\neq\varphi\in\mathcal H$, assume there is
$0\neq\varphi_{0}\in\mathcal H$ such that
\begin{equation*}
0=\langle\varphi_{0},\alpha_{ik}^{\star}\varphi\rangle=
c_{ik}\langle\varphi_{0},\psi_{k}\rangle\langle\psi_{i},\varphi\rangle
\end{equation*}
holds for all $i$ and $k$. Since by assumption $c_{ik}>0$ this
contradicts the fact that the $\psi_{i}$ form an orthonormal basis.
So this $\phi_{\underline{\alpha}}$ satisfies all conditions of
Corollary 4.1. The resulting Perron-Frobenius vector is given as the
density matrix
\begin{equation}
\label{rho}
\rho=\sum_{i}\rho_{i}|\psi_{i}\rangle\langle\psi_{i}|,\,\rho_{i}>0,
\sum_{i}\rho_{i}=1,
\end{equation}
where the $\rho_{i}$ form the components of the Perron-Frobenius vector with
eigenvalue 1 for the matrix $c_{ik}^{2}$, i.e. 
$\sum_{k}c_{ik}^{2}\rho_{k} =\rho_{i}$. As a special case 
consider the choice $c_{ik}=c_{i}>0$
for all $k$ with $\sum_{i}c_{i}^{2}=1$. Then $\rho_{i}=c_{i}^{2}$.

This discussion also gives the following converse to Corollary 4.1,
which in its
standard formulation forms one of the
starting points for Monte Carlo simulations in lattice models of
statistical mechanics and (euclidean) quantum field theory including
lattice gauge theories.
\begin{theo}
Given an arbitrary density matrix $\rho >0$, there is a
completely positive, trace preserving, compact and ergodic map $\phi$
in $\mathcal J_{1}$
for which $\rho$ is the resulting Perron-Frobenius vector. In addition
$\phi$ may be chosen to be idempotent (i.e. $\phi$ satisfies the
characteristic equation $\phi\circ (1-\phi)=0$) and
$\sigma_{1}(\phi)=\{0,1\}$.
\end{theo}

\begin{proof}Take the $\psi_{i}$'s to form an orthonormal basis in 
which $\rho$ is
diagonal, i.e. such that $\rho$ takes the form (\ref{rho}) and choose
$\phi=\phi_{\underline{\alpha}}$ with $\underline{\alpha}$ as above with
$c_{ik}=\rho_{i}^{1/2}$. Then we have $\phi(A)=\tr A\cdot\rho$ for all
$A\in \mathcal J {1}$, such that $\phi$ is idempotent. Also the kernel of $\phi$
consists of all elements which may be written in the form
$A-\tr A\cdot \rho$. Moreover let $\lambda\neq 0,1$. Then for given
$A^{\prime}$ the equation
$(\lambda-\phi)(A)=A^{\prime}$ has a solution in $A$ of the form
\begin{equation*}
A=\frac{1}{\lambda(1-\lambda)}\tr A^{\prime}\cdot\rho +
\frac{1}{\lambda}A^{\prime}.
\end{equation*}
\end{proof}

As in the corresponding ( finite dimensional) standard context $\phi$
with prescribed Perron-Frobenius vector is of course not unique.
We provide another example which in spirit is much closer to the usual
``local'' up-grading procedure in Monte Carlo simulations.
Set $p_{\nu,i}=\rho_{\nu+i}/(\rho_{i+1}+\rho_{i}+\rho_{i-1})$
for $\nu =0,\pm 1$ and with the convention $\rho_{0}=0$.
Then $\sum_{\nu =0,\pm 1}p_{\nu,i}=1$ for all $i$ and $p_{\nu,i}>0$ for all
$\nu$ and $i$ unless $\nu=-1$ and $i=1$. Set
\begin{equation*}
T_{\nu,i}=p_{\nu,i}^{1/2}\:|\psi_{\nu+i}\rangle\langle\psi_{i}|
\end{equation*}
and define $\phi_{T}$ as
\begin{equation*}
\phi_{T}(A)=\sum_{\nu,i}T_{\nu,i}AT_{\nu,i}^{\star}.
\end{equation*}
An easy calculation shows that $\phi_{T}$ is completely positive,
trace preserving, compact and ergodic.
\\
\\
\textbf{Acknowledgments} The author has profited from discussions
with D. Buchholz, B. K\"{u}mmerer, M.B. Ruskai and E. St{\o}rmer.

\end{document}